# Classical Complexity of Unitary Transformations

A. Kaltchenko
*Wilfrid Laurier University, Waterloo, Ontario, Canada, E-mail address: akaltchenko@wlu.ca*

*Abstract*:
We discuss a classical complexity of finite-dimensional unitary transformation, which can been seen as a computable approximation of descriptional complexity of a unitary transformation acting on a set of qubits.

*Keywords*:
Information theory; quantum; Kolmogorov complexity; quantum computing; quantum information; unitary.

**Introduction and notation**

Kolmogorov complexity[1] of a (classical) string or, more generally, of a classical finite object, is defined as the shortest effective binary description of that string or object. In computer science[4] and information theory[5], Kolmogorov complexity is commonly known as the shortest binary program p, which runs on a Turing machine.

Let $\mathcal{H}$ be a 2-dimensional complex vector space (Hilbert space). A qubit is described by a unit vector in $\mathcal{H}$. Then, for any integer n, the state of n qubits corresponds to a unit vector in n-folded space $\mathcal{H}^{\otimes n}$. Let $U$ be a unitary transformation on $\mathcal{H}^{\otimes n}$, represented by a $2^n$ x $2^n$ unitary matrix. It preserves inner product and has a property: $U^\dagger U = I$, $U^\dagger = U^{-1}$.

We note that any unitary transformation $U$ can be implemented as a quantum computation on Quantum Turing Machine as well as using quantum logic circuits. For a comprehensive introduction to quantum computing with elements of linear algebra and quantum mechanics, see, for example, [6].

We will use low case Latin letters to denote vectors in $\mathcal{H}^{\otimes n}$, that is $|x\rangle$, $|y\rangle$, etc.

Let $\mathcal{V}$ denote a $2^{2n}$-dimensional vector space of linear operators acting on $\mathcal{H}^{\otimes n}$.
For the vector space $\mathcal{V}$, we define a Hilbert-Schmidt inner product:
$\langle A,B\rangle_{HS} = Tr\left[A^\dagger B\right]$ (1).

Let $\left\|\langle A,B\rangle_{HS}\right\|$ denote the absolute value (modulus) of the complex number value of $\langle A,B\rangle_{HS}$.

Let $|v_i\rangle$ be an orthonormal basis set in $\mathcal{H}^{\otimes n}$. Define $|w_i\rangle \triangleq U|v_i\rangle$. Thus, $|w_i\rangle$ is also an orthonormal basis set, since unitary operators preserve inner products. Then, we have

$U \equiv \sum_i |w_i\rangle\langle v_i|$ (2).

Now we can give an operational meaning to the descriptional complexity of $U$ as follows. Given an orthonormal basis set $|v_i\rangle$ of $\mathcal{H}^{\otimes n}$, for example, a computational basis, any unitary transformation $U$ acting on $\mathcal{H}^{\otimes n}$ can be identified with- and represented by an ordered orthonormal set $\left| w_i \right\rangle$ Thus, to compute (that is, to describe) a unitary transformation *U*, it suffices to compute the corresponding orthonormal set $\left| w_i \right\rangle$.

## Main result

The Kolmogorov complexity K(|x>) of a pure quantum state |x> of n qubits was defined in [2]. It was shown in [2] that K(|x>) ≤ 2n +O(1). Note that K(|x>) is a classical description of a quantum state given by a classical binary program as opposed to a different, "entirely quantum" notion[3] of Quantum Kolmogorov complexity given by a sequence of quantum bits.

In view of the result[2], a straightforward, brute-force approach would be to encode each of $2^n$ vectors $\left| w_i \right\rangle$ separately: 2n bits per each $\left| w_i \right\rangle$, so $(2n)\cdot 2^n$ bits in total. In other words, the length of the binary encoding (which we also call a binary program p) would be $(2n)\cdot 2^n$. However, it is inconsistent with the Kolmogorov complexity of transformation of binary strings, which is a permutation or 1-to-1 mapping of a binary string of length n to itself. The descriptional complexity of such a mapping is clearly bounded by 2n.

We use a different approach and encode U as the set of $2^n$ vectors $\left| w_i \right\rangle$ *collectively* and need 4n + O(1) bits as explained further in this paper.

We use a self-delimiting Turing machine, with a binary program p, similar to the one described in [2,4], with the main difference that it will output not a sequence of qubits, but a representation of *U*. In general, a representation of *U* can be of a different form, for example, an ordered vector set $\left| w_i \right\rangle$.

For a unitary *U*, we define the output Q(*p*) of a Turing machine Q with binary program *p* as the representation of *U* given by the vector set $\left| w_i \right\rangle$. We understand that requiring the *exact* representation would result in infinite or incomputable Q(*p*) for some *U*. For such *U*, we make Q(*p*) to compute an approximation of *U*, which we denote by $U_{appr.}$ as stated in the definition below:

In the spirit of [2], we define the Kolmogorov complexity of unitary transformation *U* as follows:

*Definition 1*: $$K(U) \triangleq \min_p \left\{ l(p) + \left\lceil -\log \left\| \left\langle U_{appr}, U \right\rangle \right\|_{HS}^2 \right\rceil : Q(p) = U_{appr} \right\} \quad (3),$$

where *l(p)* is the number of bits in the program *p*.

*Definition 2*: Unitary transformations *U* is called "*directly computable*" if there is a program p such that $Q(p) = U$. Similarly, $U_{appr}$ is called the directly computable part of *U*.

Note that $U_{appr}$, an approximation of *U*, is a unitary operator, which is produced by the computation *Q(p)*, and therefore, given *Q*, completely determined by *p*. Thus, we obtain the minimum of the right-hand side of the equality by minimizing over *p* only.

The value of $\left\| \left\langle U_{appr}, U \right\rangle \right\|_{HS}^2$ can be seen as the value of the probability measure, which gives the probality of vector $U$, given vector $U_{appr}$.

Thus, Quantum Kolmogorov complexity $K(U)$ is the sum of two terms: the first term is the integral length of a Kolmogorov binary program, which directly encodes $U_{appr}$, and the second term, the min-log quasi-probability term, corresponds to the length $\left\lceil -\log \left\| \left\langle U_{appr}, U \right\rangle \right\|_{HS}^2 \right\rceil$ of the Shannon-Fano codeword for $U$ associated with that probability distribution, see for example [5], and is thus also expressed in an integral number of bits.

Throughout the paper, we will omit the "quasi" prefix in "(quasi)probability" and will use term "probability".

To describe Shannon–Fano encoding, we see unitary operators as vectors in the $2^{2n}$-dimensional vector space $\mathcal{V}$. Given vector $U_{appr}$ and an orthonormal basis of $\mathcal{V}$ with $U$ as one of the basis vectors, we encode $U$ using the Shannon–Fano prefix code as follows.

Let $B = \left\{ \left| e_1 \right\rangle, \ldots, \left| e_{2^{2n}} \right\rangle \right\}$ be an orthonormal basis of the vector space $\mathcal{V}$.

Then, we have

$$\sum_{i=1}^{2^{2n}} \left\| \left\langle e_i, U_{appr} \right\rangle \right\|_{HS}^2 = 1 \quad (4).$$

If we let $U$ be one of the basis vectors, then we can consider vector $U_{appr}$ as a random variable that assumes the value of vector $U$ with probability $\left\| \left\langle U_{appr}, U \right\rangle \right\|_{HS}^2$.

The Shannon–Fano codeword for $U$ in the probabilistic ensemble

$$\left\{ \left( \left| e_1 \right\rangle, \left\| \left\langle e_1, U_{appr} \right\rangle \right\|_{HS}^2 \right), \left( \left| e_2 \right\rangle, \left\| \left\langle e_2, U_{appr} \right\rangle \right\|_{HS}^2 \right), \left( \left| e_3 \right\rangle, \left\| \left\langle e_3, U_{appr} \right\rangle \right\|_{HS}^2 \right), \ldots, \left( \left| e_{2^{2n}} \right\rangle, \left\| \left\langle e_{2^{2n}}, U_{appr} \right\rangle \right\|_{HS}^2 \right) \right\} \quad (5)$$

is based on the probability $\left\| \left\langle U_{appr}, U \right\rangle \right\|_{HS}^2$ of $U$ given $U_{appr}$ and has length $\left\lceil -\log \left\| \left\langle U_{appr}, U \right\rangle \right\|_{HS}^2 \right\rceil$.

Essentially, the Shannon–Fano codeword for $U$ encodes the index $i$ in the basis vector set $B$, where $i$ is such that $\left| e_i \right\rangle = U$.

Note that our codeword lengths $l_i$ satisfy the Kraft inequality[4,5] ensuring that our encoding is uniquely decodable:

$$\sum_{i=1}^{2^{2n}} 2^{-l_i} \leq 1 \quad (6),$$

where $l_i = \left\lceil -\log \left\| \left\langle \mathrm{e}_i, U_{appr} \right\rangle \right\|_{HS}^2 \right\rceil$.

With a canonical Gram–Schmidt process of constructing an orthonormal basis from a given basis vector, we can choose $B$ such that $K(B) = \min_i \left\{ K\left( \left| e_i \right\rangle \right) \right\}$.

Essentially, we have extended the Kolmogorov complexity[2] for a quantum state, given by a vector in $\mathcal{H}^{\otimes n}$, to that for a vector in the space of linear operators if we view *U* as a vector in the space of linear operators acting on $\mathcal{H}^{\otimes n}$.

Loosely speaking, K(*U*) can bee seen as the number of bits, which is required to compute (in other words, to describe) the ordered vector set $|w_i\rangle$ for a unitary transformation *U*.

*Theorem 1*: We obtain an upper bound: K(*U*) $\leq$ 4n + O(1).

*Proof:*
The idea of the proof is as follows:
To encode *U*, we need 2n +O(1) bits to encode a computational basis as well as 2n +O(1) bits to encode coordinates (projections) in that basis. Thus, it takes 2n +O(1) + 2n +O(1) = 4n +O(1) in total. Such a result can be seen as an extension of Theorem 3 (Upper Bound) in [2] to the vector space of linear operators acting on $\mathcal{H}^{\otimes n}$.

In the proof, we see unitary operators as vectors in the $2^{2n}$-dimensional vector space $\mathcal{V}$ of linear operators acting on $\mathcal{H}^{\otimes n}$.
Let *p* be a binary program to construct a basis state $|e_i\rangle$ of $\mathcal{V}$. We need 2n bits to enumerate the set of $2^{2n}$ basis vectors, therefore, we have $l(p) \leq 2n+O(1)$, where $l(p)$ is the length of the binary program *p*.

For every vector *U* in the $2^{2n}$-dimensional vector space $\mathcal{V}$ with basis vectors $|e_1\rangle, |e_2\rangle, \ldots, |e_{2^{2n}}\rangle$, we have $\sum_{i=1}^{2^{2n}} \left\| \langle e_i | U \rangle \right\|^2 = 1$ . So, there is an *i* such that $\left\| \langle e_i | U \rangle \right\|^2 \geq 1/2^{2n}$ . Note that, for directly computable states, we archive the equality $\left\| \langle e_i | U \rangle \right\|^2 = 1$ .
If we were to relax optimization in (3) and have $U_{appr} = |e_i\rangle$, then the complexity would be bounded by $l(p) + \log 2^{2n}$ . Thus, $K(U) \leq l(p) + \log 2^{2n} \leq 4n + O(1)$ .

**Discussion**

The Kolmogorov complexity of a unitary transformation on $\mathcal{H}^{\otimes n}$ is closely related to the Quantum Kolmogorov complexity[2] of a quantum states in $\mathcal{H}^{\otimes n}$ as we see in the following example:

Let $|0\otimes\ldots\otimes 0>$ be a vector in an orthonormal computational basis in $\mathcal{H}^{\otimes n}$.
Let $|y> = U\,|0\otimes\ldots\otimes 0>$.

Then, we have:
$K(|y>) \leq K(U) + O(1)$, where $K(|y>)$ is the Quantum Kolmogorov complexity[2] of the quantum state $|y>$.

## Conclusion

We point out that the exact description (that is computing) of an arbitrary finite-dimensional unitary transformation *U* is generally incomputable or infinite. So, in the most general case, K(*U*) is the length of the binary program which *approximates U*. Nevertheless, it fits nicely with *discrete* Turing machines as well as with quantum circuits and quantum algorithms with a *discrete* set of quantum gates. Thus, K(*U*) reflects the (classical) computational complexity of a quantum circuit in the Kolmogorov sense and provides a tool and theoretical framework for the analysis of quantum circuit complexity.

*Future work*:

Given a computational basis of $\mathcal{H}^{\otimes n}$, study how the Kolmogorov complexity of a unitary transformation *U* is related to the Kolmogorov complexity of quantum states produced by *U*.